\documentclass[reprint,aps,twocolumn,showpacs,superscriptaddress,groupedaddress,nofootinbib]{revtex4-2}
\usepackage{graphicx}  
\usepackage{tabularx}
\usepackage{booktabs}
\usepackage{dcolumn}   
\usepackage{bm}        
\usepackage{amssymb}   
\usepackage{amsmath}   
\usepackage{multirow}
\usepackage{hyperref}
\usepackage{braket}
\usepackage{subfigure}
\usepackage{array}
\usepackage[math]{cellspace}
\usepackage[utf8]{inputenc}
\usepackage{cleveref}
\usepackage{algorithm}
\usepackage{algpseudocode}
\usepackage{physics}
\algrenewcommand\algorithmicrequire{\textbf{Input:}}
\algrenewcommand\algorithmicensure{\textbf{Output:}}

\usepackage{color}
\usepackage[normalem]{ulem}

\usepackage{url}
\hypersetup{
bookmarksnumbered,
colorlinks=true,
linkcolor=magenta,
urlcolor=magenta,
citecolor=blue
} 
\begin{document}
\hspace{5.2in} 

\title{Strong $CP$ problem in the quantum rotor}

\date{\today}

\author{D. Albandea}
\email{david.albandea@uv.es}
\affiliation{Instituto de Física Corpuscular (CSIC-UVEG), Edificio Institutos Investigaci\'on,
Apartado de Correos 22085, E-46071 Valencia, Spain}
\author{G. Catumba}
\email{gtelo@ific.uv.es}
\affiliation{Instituto de Física Corpuscular (CSIC-UVEG), Edificio Institutos Investigaci\'on,
Apartado de Correos 22085, E-46071 Valencia, Spain}
\author{A. Ramos}
\email{alberto.ramos@ific.uv.es}
\affiliation{Instituto de Física Corpuscular (CSIC-UVEG), Edificio Institutos Investigaci\'on,
Apartado de Correos 22085, E-46071 Valencia, Spain}

\begin{abstract}
    Recent studies have claimed that the strong $CP$ problem does not occur in
    QCD, proposing a new order of limits in volume and topological sectors when
    studying observables on the lattice.  In order to shed light on this issue,
    we study the effect of the topological $\theta$-term on a simple quantum
    mechanical rotor that allows a lattice description. The topological
    susceptibility and the $\theta$-dependence of the energy spectrum are both
    computed using local lattice correlation functions. The sign problem is
    overcome by considering Taylor expansions in $\theta$ exploiting automatic
    differentiation methods for Monte Carlo processes. Our findings confirm the
    conventional wisdom on the strong $CP$ problem.
\end{abstract}
\maketitle

\section{Introduction}
\label{sec:introduction}

The strong $CP$ problem remains one of the puzzles of the Standard Model: Why
do strong interactions conserve $CP$?  In principle, the Lagrangian of quantum
chromodynamics (QCD) admits a renormalizable gauge invariant $\theta$-term,
\begin{equation}
  \label{eq:Ltheta}
  \delta \mathcal L_\theta = \frac{\theta}{16\pi^2} {\rm tr} \left(
  F_{\mu\nu}\tilde F^{\mu\nu} \right)\,,
\end{equation}
where $F_{\mu\nu} = \partial_\mu A_\nu - \partial_\nu A_\mu + [A_\mu,A_\nu]$ is
the field strength and $A_\mu$ the gauge connection.  While the presence of such
a term would break $CP$ symmetry in QCD, experimental measurements of the electric
dipole moment of the neutron constrain the coupling to be
$|\theta| \lesssim 10^{-10}$ \cite{Abel:2020pzs,RevModPhys.91.015001}.

Many solutions have been proposed over the years. The existence of a massless
quark would make the $\theta$ phase unphysical, but recent lattice simulations
clearly disfavor such a solution
\cite{FlavourLatticeAveragingGroupFLAG:2021npn}.  Additionally, various
alternatives beyond the Standard Model, such as a Peccei-Quinn symmetry
\cite{PhysRevD.16.1791} and Nelson--Barr-type models
\cite{Nelson:1983zb,PhysRevLett.53.329}, have been explored.

References~\cite{ai_absence_2021,ai_consequences_2021} claimed that
the effect of a such a $\theta$-term, even if present, would not lead to
observable consequences. These works argued that when computing correlation
functions, the infinite volume limit should be taken \emph{before} the summation
over topological sectors. The consequence of such an order of limits (infinite
volume at each fixed topological charge) would be the absence of
$\theta$-dependence from observables, and particularly from the energy spectrum
of the theory.

We claim that, in fact, the order of limits is only important if one insists on
computing \emph{global} observables, i.e. those computed as an integral over the
whole Euclidean space. Determining these observables requires extreme care with
the role of boundary conditions and in taking the infinite volume limit. On the
other hand they also represent quite unphysical setups, since we do not need to
know the boundary conditions of the universe to measure the topological
susceptibility, the proton mass, or the neutron electric dipole moment: All
physical quantities of interest can be extracted from local correlators. Because
of clustering, the dependence of local correlators on the boundary conditions or
the finite volume is exponentially suppressed in theories with a mass gap,
making the order of limits largely irrelevant.

In this paper, we aim to shed some light on this question by examining a simple
toy model that, nevertheless, shares some key characteristics with QCD: the
quantum rotor. We will compute the topological susceptibility from local
correlators and show that, up to finite volume corrections, the result is
independent of the choice of boundary conditions.  The analytical computations
will also be supported by numerical lattice simulations,\footnote{The code used
    for the simulations and analysis can be found in
    \url{https://github.com/dalbandea/QuantumRotorExperiments.jl} and
\url{https://igit.ific.uv.es/gtelo/qrotor}.} validating the approach used
by the lattice community to answer these questions in QCD
\cite{Shindler:2015aqa,Dragos:2019oxn,Giusti:2018cmp}.  Additionally, we will
determine the topological susceptibility using master field
simulations~\cite{Luscher:2017cjh}, where a single gauge configuration in a very
large Euclidean volume allows us to determine expectation values as volume
averages. Both local computations give the same nonzero value for the
topological susceptibility, $\chi_t$. Moreover, we will show that the spectrum
of the theory has a dependence on $\theta$.


Finally, the numerical study of this rather simple system faces several
challenging problems that are present in lattice QCD.  We will use several recent
proposals to overcome these problems, making the study of this particular model
a good test bed for many state-of-the-art lattice techniques.  First, the
issue of topology freezing
\cite{Alles:1996vn,DelDebbio:2002xa,DelDebbio:2004xh,Schaefer:2010hu} when
approaching the continuum limit is solved by the use of winding transformations
\cite{Albandea_2021}. Second, simulating the theory at a nonzero value of
$\theta$ leads to the so-called sign problem ---see
\cite{deForcrand:2009zkb,Gattringer:2016kco} and references therein.  To
overcome this, we employ three different methods, which rely on the use of imaginary
values for $\theta$ along with analytical
continuation~\cite{Bhanot:1984rx,Azcoiti:2002vk,Alles:2007br,Alles:2014tta,Panagopoulos:2011rb,DElia:2012pvq,Bonati:2015sqt,Aoki:2008gv,Bonati:2016tvi,Bonanno:2023hhp,bonanno2024thetadependence}:
On the one hand, we will extract the $\theta$ dependence by a fit to data from
several simulations at different imaginary values of $\theta$; on the other
hand, we will explore two recent proposals that allow us to extract series
expansions in $\theta$ using automatic differentiation for Monte Carlo
processes~\cite{Catumba:2023ulz}.

The paper is organized as follows: In Sec.~\ref{sec:quant-mech-rotor} we introduce
the quantum rotor, along with the action and topological charge definitions in
the continuum and on the lattice, including two different choices of
discretization; in Sec.~\ref{sec:chi_t-from-local} we study the
order of limits argued in Refs.~\cite{ai_absence_2021,ai_consequences_2021}  and
the conventional one using both global and local observables, where the latter
are computed with lattice simulations; in Sec.~\ref{sec:theta-dependence} we obtain
the continuum $\theta$-dependence of the energy spectrum from lattice
simulations using different boundary conditions---this computation requires
overcoming the sign problem, whose proposed solutions are introduced in
Sec.~\ref{sec:signproblem}.

\section{Quantum rotor}
\label{sec:quant-mech-rotor}

The quantum rotor describes a free particle of mass $m$ on a ring of radius $R$.
The system admits a $\theta$-term and can be formulated at finite temperature
$\beta = 1 / T$ as a path integral with partition function
\cite{Fjeldso:1987wi,bietenholz_perfect_1997}
\begin{align}
    \label{eq:qr-partition-fct}
    Z(\theta) = \int \mathcal{D}\phi\, e^{-S(\phi) + i\theta Q},
\end{align}
where the action reads
\begin{equation}
    \label{eq:action-qr}
    S(\phi) = \frac{I}{2} \int_{0}^{T} dt\, \dot \phi(t)^2 \,,
\end{equation}
with $\phi(0) =  \phi(T)$ and $-\pi < \phi(t)\le \pi$, and where we have defined
the moment of inertia $I = mR^2$. Note that using periodic boundary conditions
leads to the quantization of the topological charge,
\begin{equation}
    \label{eq:top-qr}
  Q = \frac{1}{2\pi} \int_0^T dt\, \dot\phi(t)\, \in \mathbb{Z}.
\end{equation}

To study the system on a lattice we divide the time extent $T \equiv \hat{T} a$
into a lattice of $\hat{T}$ points with spacing $a$, and express the moment of
inertia measured in units of this lattice spacing as $\hat{I}=I / a$. There are
several possibilities for the discretization of the lattice action, with the
only condition being that the continuum action and topological charge in
Eqs.~(\ref{eq:action-qr}) and (\ref{eq:top-qr}) are recovered in the limit
$\hat{I} \to \infty$ with
\begin{align}
    \frac{T}{I} = \frac{\hat{T}}{\hat{I}} = \text{constant}.
\end{align}
Particularly, we will use the so-called classical perfect action
\cite{Manton:1980ts},
\begin{equation}
    \label{eq:cp-s-qr}
  S_{\rm cp}(\phi) = \frac{\hat I}{2}\sum_{t=0}^{\hat{T}-1}
  ((\phi_ {t+1}-\phi_t)\bmod 2\pi)^2\\,
\end{equation}
as well as the standard action,
\begin{equation}
    \label{eq:st-s-qr}
  S_{\rm st}(\phi) = \frac{\hat I}{2}\sum_{t=0}^{\hat{T}-1}
  (1-\cos(\phi_ {t+1}-\phi_t))\,.
\end{equation}
One can show that the classical perfect action has smaller lattice artifacts and
therefore a better behavior as one approaches the continuum limit
\cite{bietenholz_perfect_1997}. However, the standard discretization is useful
when using sampling algorithms that require the derivative of the action with
respect to the field to be defined at all points of its domain. Particularly, we
use the hybrid Monte Carlo (HMC)~\cite{DUANE1987216,brida_smd-based_2017}
algorithm to overcome the sign problem when studying the $\theta$ dependence of
the spectrum of the theory.

Analogously, we define the classical perfect topological charge,
\begin{equation}
    \label{eq:cp-Q-qr}
  Q_{\rm cp} = \frac{1}{2\pi} \sum_{t=0}^{\hat{T}-1}
  ((\phi_ {t+1}-\phi_t)\bmod 2\pi),
\end{equation}
and the standard topological charge,
\begin{equation}
    \label{eq:st-Q-qr}
  Q_{\rm st} = \frac{1}{2\pi} \sum_{t=0}^{\hat{T}-1}
  \sin(\phi_ {t+1}-\phi_t).
\end{equation}
Note that with periodic boundary conditions the classical perfect topological
charge is exactly an integer, while the standard one is only an integer in the
continuum limit.

\section{Order of limits of infinite volume and topological sectors}
\label{sec:chi_t-from-local}

For theories with topology, the computation of an observable can be decomposed
as a sum of contributions from the different topological sectors,
\begin{equation}
  \expval{\mathcal O} = \sum_{Q} \expval{\mathcal O}_{Q} p(Q),
\end{equation}
where $p(Q)$ is the probability density of the topological sector with charge
$Q$, and $\left< \dots \right>_{Q}$ denotes the expectation value at fixed
topological sector $Q$.
While this equation is general, it requires the topological charge $Q$ to
be quantized.  For gauge quantum field theories, the usual reasoning for the
quantization of the topological charge arises from the requirement of finite
saddle solutions for the action,\footnote{Not only does the requirement of finite
action solutions come from the use of a semiclassical approximation, but
the topological quantization can also be obtained for a finite volume with
appropriate boundary conditions \cite{Coleman_1985}.} which constrains the gauge
configurations at infinity to be pure gauge (i.e. a gauge transformation of
zero).

In practice, lattice calculations are restricted to finite Euclidean volumes.
This, together with the above reasoning, led the authors of
Refs.~\cite{ai_absence_2021,ai_consequences_2021} to challenge the conventional
order of limits to extract observables at infinite volume. Namely, they proposed
computing observable quantities as
\begin{equation}
  \expval{\mathcal O} = \lim_{N\rightarrow\infty} \lim_{T\rightarrow\infty}\sum_{\abs{Q}<N} \expval{\mathcal O}_{Q} p(Q),
\end{equation}
such that the volume is taken to infinity before the contributions from all the
topological sectors are summed.

This claim can be verified in the quantum rotor, which can be trivially solved
using the quantum mechanical formalism~\cite{bietenholz_perfect_1997}. The energy
levels of the system read
\begin{equation}
  \label{eq:spectrum continuum}
  E_n = \frac{1}{2I} \left( n - \frac{\theta}{2\pi} \right)^2, \quad n\in\mathbb
  Z\,.
\end{equation}
In Euclidean time with periodic boundary conditions, we have the thermal
partition function
\begin{equation}
  \label{eq:Z}
  Z(\theta) = \sum_{n\in\mathbb Z} e^{- T E_{n} } = \sum_{n\in\mathbb Z} e^{- \frac{T}{2I}\left( n-
  \frac{\theta}{2\pi} \right)^2 },
\end{equation}
from which we can obtain the second moment of the topological charge at
$\theta =0$,
\begin{align}
  \langle Q^2 \rangle &= - \frac{\partial^2 \log Z(\theta)}{\partial
  \theta^2}\Big|_{\theta=0}  \nonumber \\
  &=\frac{T}{4I\pi^2} \left( 1 -  \frac{T}{I}\frac{\sum_nn^2 e^{- \frac{T}{2I}n^2}}{\sum_n
  e^{- \frac{T}{2I}n^2}} \right)\,.
  \label{eq:Q2 sum}
\end{align}
Note that for large $T$ we have
\begin{equation}
  \frac{\langle Q^2 \rangle}{T} = \frac{1}{4I\pi^2} \left[ 1 - 2\frac{T}{I}\left(
  e^{- \frac{T}{2I}} - e^{- \frac{T}{I}}\right) + \dots \right],
\end{equation}
i.e., a nonzero value with exponentially small corrections due to the finite
size of the system. This leads to the usual result for the topological
susceptibility,
\begin{equation}
  \label{eq:chicorrect}
  \chi_t = \lim_{T\to\infty} \frac{ \langle Q^2 \rangle }{T} = \frac{1}{4I\pi^2}\,.
\end{equation}

We can, instead, attempt to take the infinite $T$ limit before summing over
topological sectors. To do this, we determine the probability distribution
of each topological sector of charge $Q$. Considering $\theta = 0$ for
simplicity we have
\begin{align}
    p(Q) =& \frac{1}{2\pi Z(0)} \int_{-\pi}^{\pi} d\theta\,Z(\theta)e^{-i\theta Q}
    \nonumber \\
    =& \frac{1}{Z(0)}\sqrt{ \frac{2I\pi}{T} } \exp \left( - \frac{2I\pi^2}{T}
    Q^2 \right).
\end{align}
According to Ref.~\cite{ai_absence_2021}, the correct order of limits should now read
\begin{align}
  \chi_t &= \lim_{N\to\infty} \lim_{T\to\infty} \frac{1}{T} \sum_{\abs{Q}<N} Q^2 \,p(Q)\,\nonumber\\
  &= \lim_{N\to\infty} \lim_{T\to\infty} \frac{1}{T}\frac{\sum_{|Q|\leq N} Q^{2}
  \exp( -\frac{2\pi^{2}I}{T}Q^{2})}{\sum_{|Q|\leq N} \exp(
-\frac{2\pi^{2}I}{T}Q^{2})},
    \label{eq:sum sectors}
\end{align}
which is trivially zero and therefore contradicts the result in
Eq.~(\ref{eq:chicorrect}).



\begin{figure}
  \centering
  \includegraphics[width=0.49\textwidth]{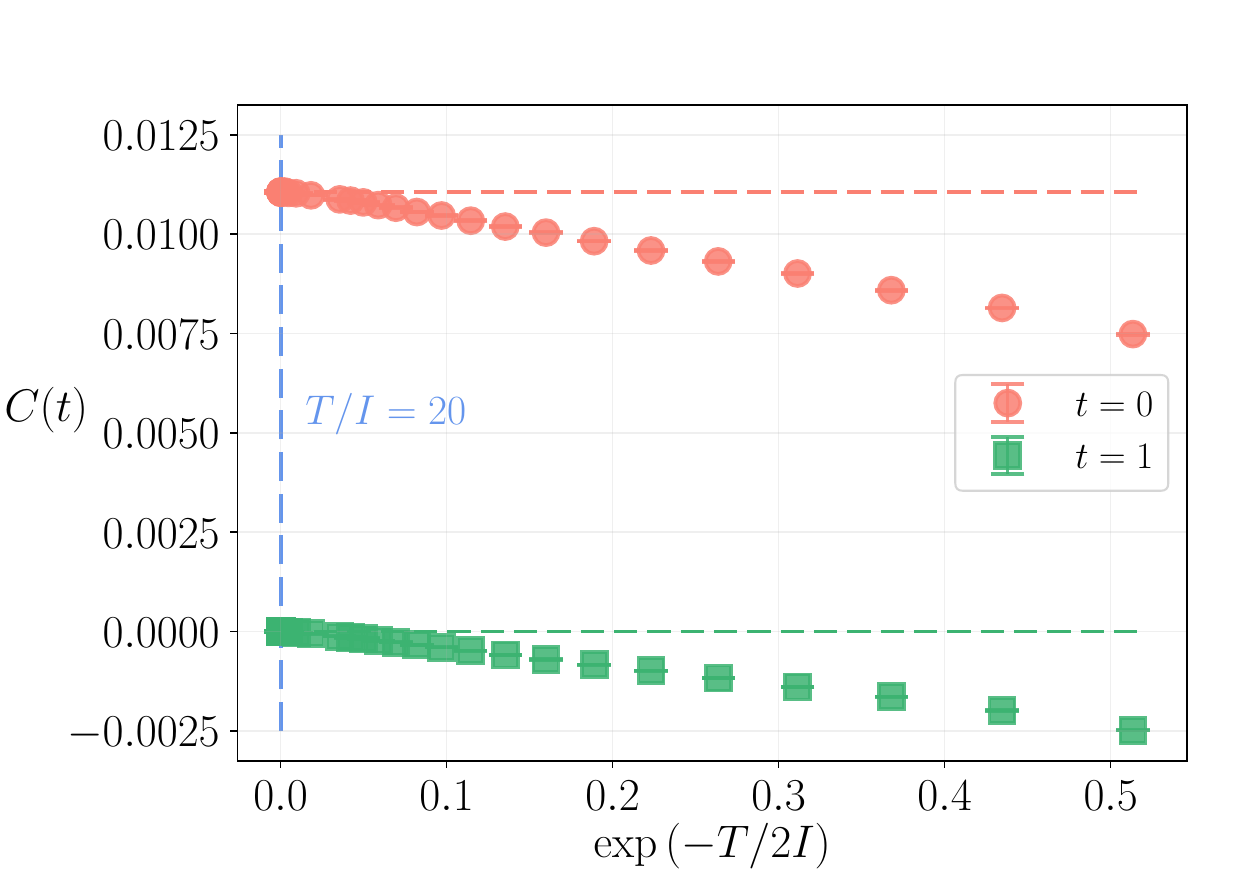}
  \caption{Quantity $C(t) \equiv \langle q(t)q(0) \rangle$ as a function of
      $\exp \left( -T / 2I \right)$ from simulations at $\hat{I}=3$
and $\hat{T}\in[4,120]$ with the standard action  with periodic boundary
conditions at $t=0$ (red circles) and $t=1$ (green squares). The horizontal dashed lines
represent the corresponding analytical results with open boundary conditions.
The vertical dashed line indicates the position of $T / I = 20$, corresponding
to the line of constant physics at which the simulations in this work are
performed.}
  \label{fig:expeffects}
\end{figure}

To shed some light on this discrepancy, we propose another approach in which
the order of limits is irrelevant. Because of locality, Euclidean two-point
functions decay exponentially at large time separations, a fact that can be
exploited to compute the topological susceptibility from local correlators. If
one defines the topological charge density as $q(t) = \frac{1}{2\pi}\dot \phi(t)$, we have
\begin{eqnarray}
  \nonumber
  \langle Q^2 \rangle &=& \sum_{t_1, t_2} \langle q(t_1)q(t_2) \rangle =
  T \sum_t \langle q(t)q(0) \rangle \\
  &=& T \sum_{t < R}\langle q(t)q(0) \rangle +
  \mathcal O(e^{-R / \xi})\,,
\end{eqnarray}
where the last sum runs over
a time extent $R$ and $\xi$ represents the correlation length of the system,
which, for the quantum rotor, reads $\xi =(E_{1}-E_{0})^{-1}=2I$.
Therefore, one can compute the topological susceptibility in a localized region
of radius $R \ll T$,
\begin{equation}
    \label{eq:chilocal}
    \chi_t = \sum_{t<R}\langle q(t)q(0) \rangle + \mathcal{O}(e^{-R / \xi}),
\end{equation}
where contributions from $t > R$ are exponentially suppressed. We make the
following observations:
\begin{enumerate}
\item Since $R \ll T$ and finite volume effects in Eq.~(\ref{eq:chilocal}) are
    $\mathcal{O}(e^{-T / \xi})$, they are even more exponentially suppressed. As
    we will see, any choice of boundary conditions must give the same result.
  
\item In Eq.~(\ref{eq:chilocal}), there is no sum over topological sectors, and, in
    fact, $\displaystyle \frac{1}{2\pi}\sum_{t<R} q(t)$ is not quantized. There
    is no order of limits to discuss.
\end{enumerate}
It is easy to compute observables in the quantum rotor using open boundary
conditions.\footnote{ This amounts to setting $\phi_{\hat{T}} = \phi_{\hat{T}-1}$
    in Eqs~(\ref{eq:cp-s-qr})--(\ref{eq:st-Q-qr}), which corresponds to
    Neumann boundary conditions in the continuum, i.e. $\dot \phi(T) = \dot
\phi(0) = 0$.} Concretely, the two-point correlation function of the topological
density, which we derive in Appendix~\ref{sec:expl-comp-with}, yields
\begin{equation}
  \label{eq:2ptobc}
  \langle q(t_1)q(t_2) \rangle_{\text{cp, OBC}} = \delta_{t_1,t_2}
  \left[ \frac{1}{4I\pi^2} + \mathcal{O}(I^{-2}) \right]\,.
\end{equation}
In this particular case, finite time-extent effects are completely absent and
one can just set $R=1$ in Eq.~(\ref{eq:chilocal}). In the continuum limit, we then
have
\begin{align}
    \lim_{I \to \infty} I \chi_{t} = \frac{1}{4\pi^2},
\end{align}
which coincides with the result in Eq.~(\ref{eq:chicorrect}).

\begin{figure}
  \centering
  \includegraphics[width=0.49\textwidth]{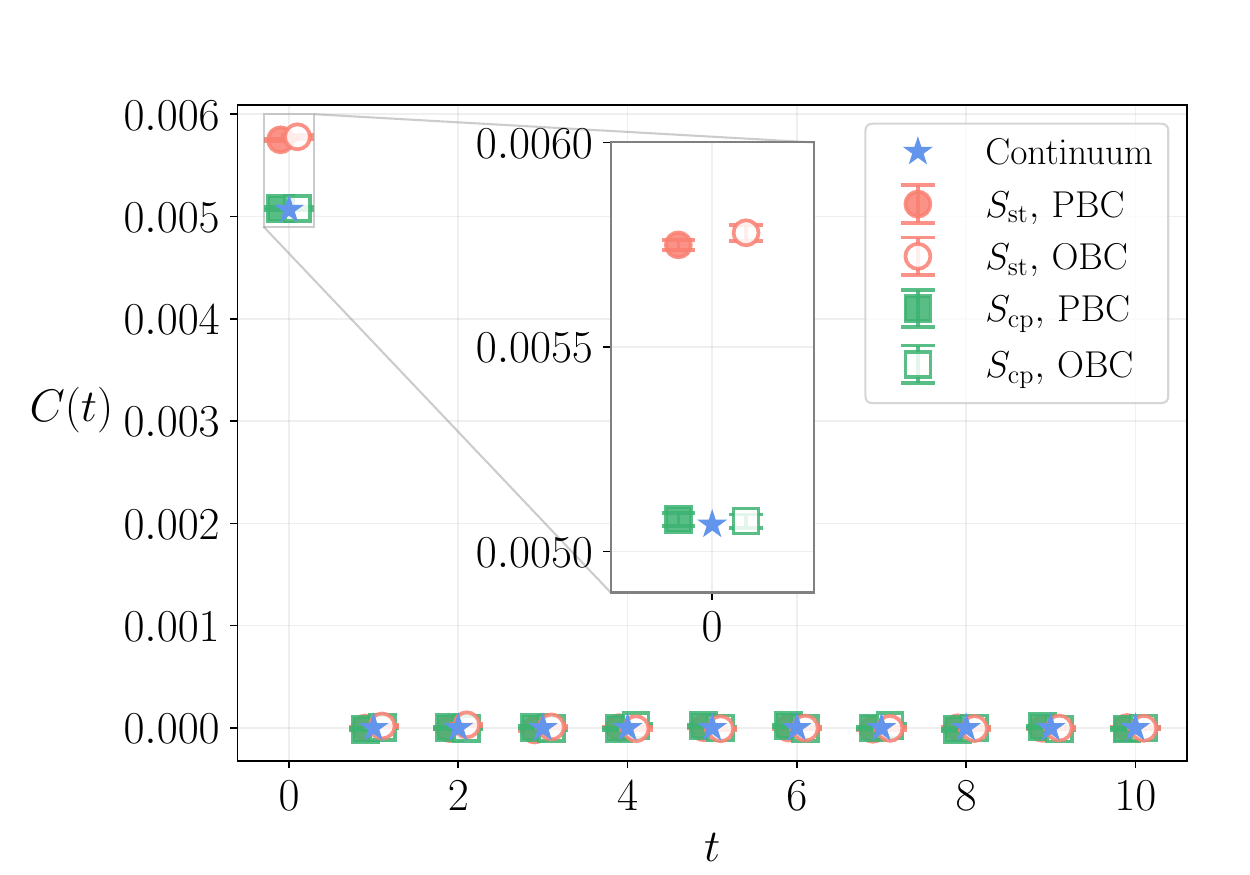}
  \caption{Quantity $C(t) \equiv \langle q(t)q(0) \rangle$ as a function of
      $t$ from simulations at $\hat{I}=5$ and $\hat{T}=100$ with the standard
      action (red circles) and classical perfect action (green squares) with
      open boundary conditions (open symbols) and periodic boundary conditions
      (filled symbols). The blue stars represent the continuum result.}
  \label{fig:contact}
\end{figure}

One can use lattice simulations to obtain the same result using periodic
boundary conditions instead. Although in this model topology freezing is already
apparent at rather small values of $\hat{I}$, we overcome it by
using a combination of standard Metropolis updates \cite{Metropolis1953} and
winding transformations \cite{Albandea_2021} (see Appendix~\ref{app:winding-trafo}).

It is worth noting that, in the case with periodic boundary conditions, the
topological two-point function should have a $t$-dependence similar to the
correlator for open
boundaries, $\langle q(t_1)q(t_2) \rangle \propto \delta_{t_1,t_2}$, up to
exponentially small finite-volume effects in $T / I$. This can be clearly seen
in Fig.~\ref{fig:expeffects}, where we show $\langle q(0)q(0) \rangle$ (red
circles) and $\langle q(1)q(0) \rangle$ (green squares) as a function
of $\exp(-T / 2I)$ from simulations with the standard action at $\hat{I}=3$ and
for different values of $\hat{T}\in[4,120]$. Both quantities
exponentially approach the corresponding analytical results with open boundary
conditions (which do not have finite-$T$ effects), depicted by the horizontal
dashed lines. The vertical dashed line corresponds to $T / I = 20$, at which
finite-$T$ effects are negligible with respect to the statistical precision of
the simulations. In Fig.~\ref{fig:contact} we additionally show the dependence
on $t$ for simulations at $\hat{I}=5$ and
$\hat{T}=100$, for the standard action (red circles) and the classical perfect
action (green squares), as well as for open (open symbols) and periodic (filled
symbols) boundary conditions. The continuum result is also shown (blue stars).

\begin{figure}
  \centering
 \includegraphics[width=0.49\textwidth]{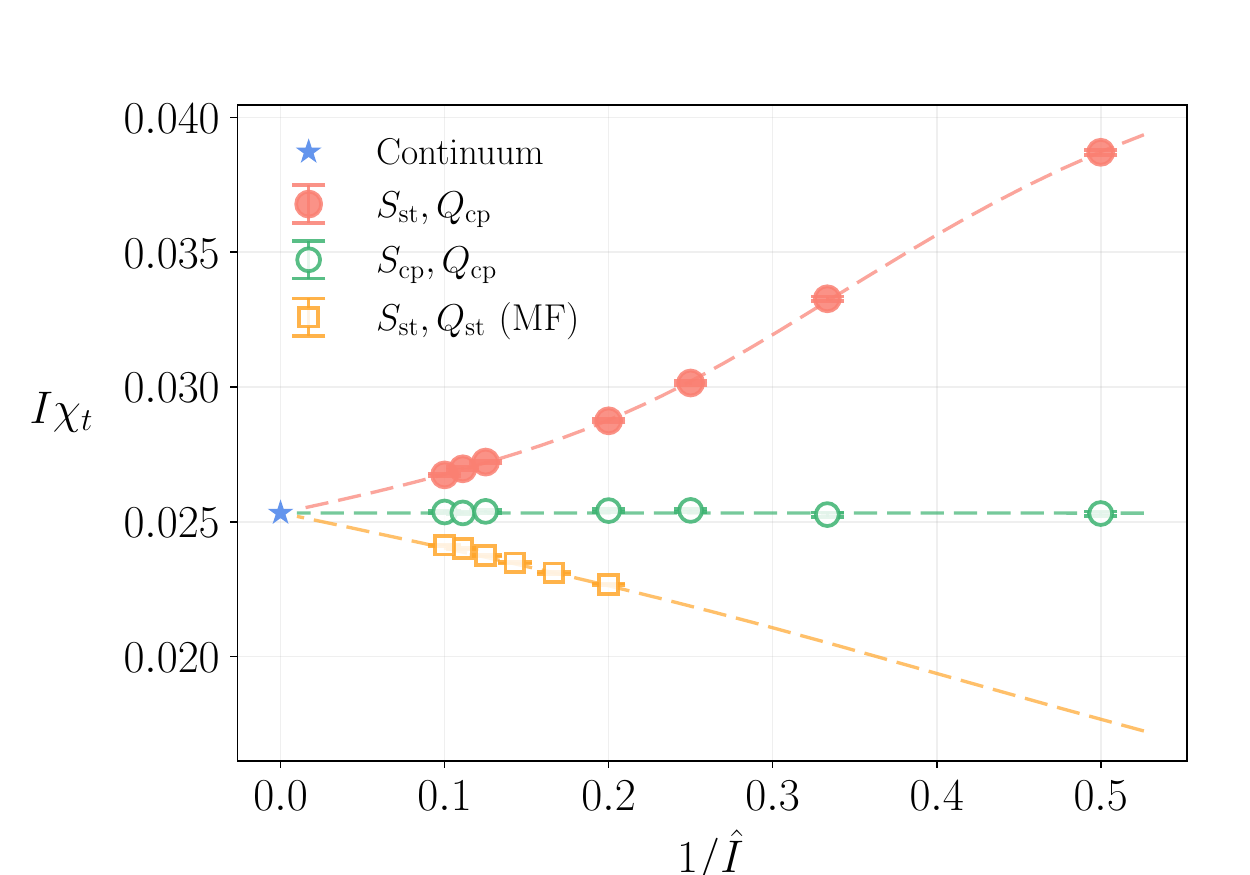}
  \caption{Continuum extrapolation of $\chi_t$ computed from $ \langle q(t)^2
      \rangle$ at various $\{\hat{T},\hat I\}$ at constant $T/I=20$ with the
      standard action and classical topological charge (red, filled circles),
      with the classical action and topological charge (green, open circles) and
      from master field simulations with the standard action and topological
      charge (orange, open squares), all with periodic boundary conditions. The
      topological charges of the master fields, from smaller to larger value of
      $\hat{I}$, are $Q=-83, -61, -233, 11, 24, -118$. The dashed lines
      represent the corresponding analytical results with open boundary
  conditions.}
  \label{fig:chit}
\end{figure}

On the other hand, the discrepancy between the standard and classical perfect
actions is a discretization effect that disappears when taking the continuum limit.
This can be seen in Fig.~\ref{fig:chit}, where we show the continuum extrapolation
of $I \chi_{t}$ with simulations at constant $T / I = 20$ for the classical
perfect action (green, open circles) and for the standard action (red, filled
circles) with periodic boundary conditions. The red and green dashed lines are
the corresponding analytic curves with open boundary conditions, and one can see
that both choices of discretization and boundary conditions give the same
nonzero results for $\hat I\rightarrow\infty$.

Finally, the orange squares come from master field simulations using the
standard action with periodic boundary conditions. Concretely, every point comes
from the volume average of the local definition of the topological
susceptibility in Eq.~(\ref{eq:chilocal}) on a single master field configuration with
$T=10^{6}$, and with the standard definition of the topological charge of
Eq.~(\ref{eq:st-Q-qr}). Although these configurations can have very large
topological charge values, the result, once extrapolated to the
continuum, matches Eq.~(\ref{eq:chicorrect}).

\begin{figure*}[!t]
  \centering
 \includegraphics[width=0.49\textwidth]{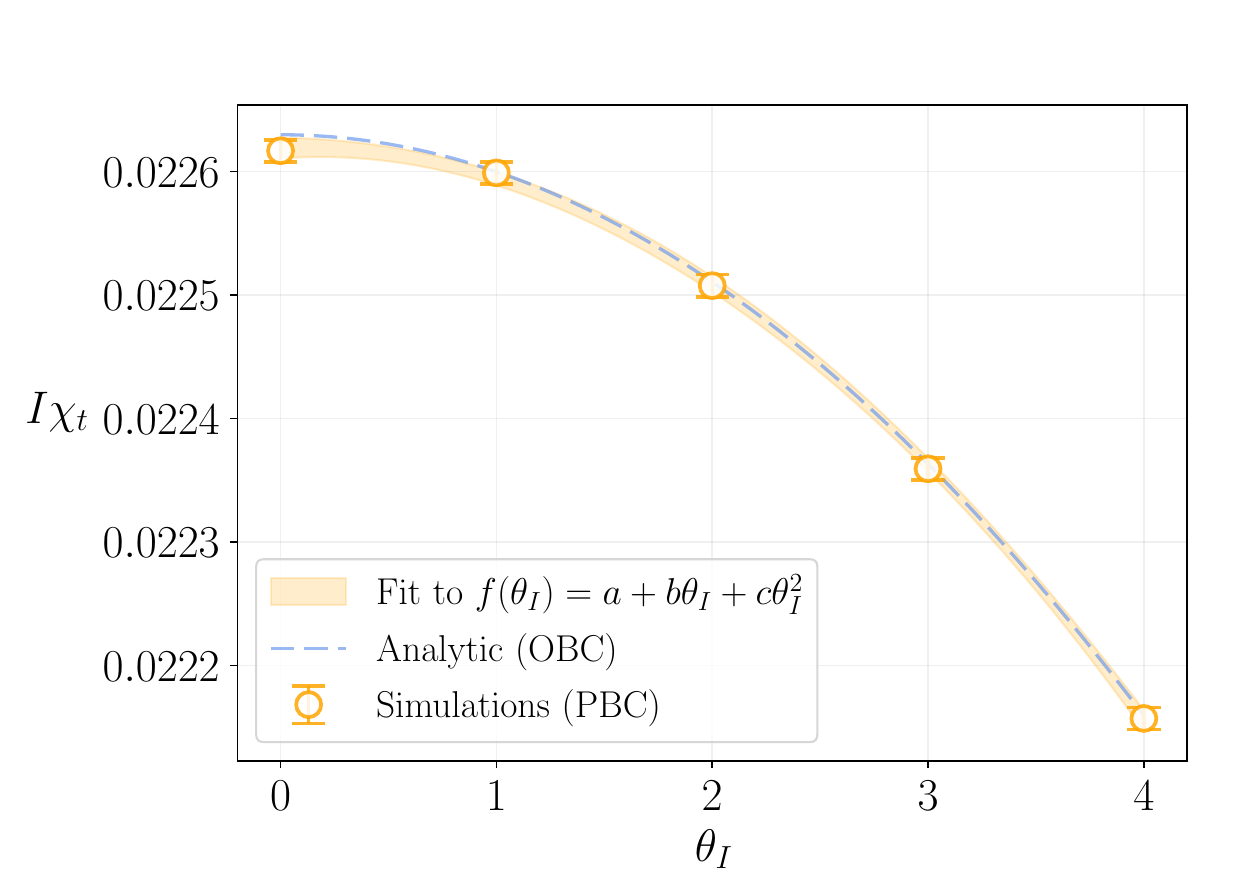}
 \includegraphics[width=0.49\textwidth]{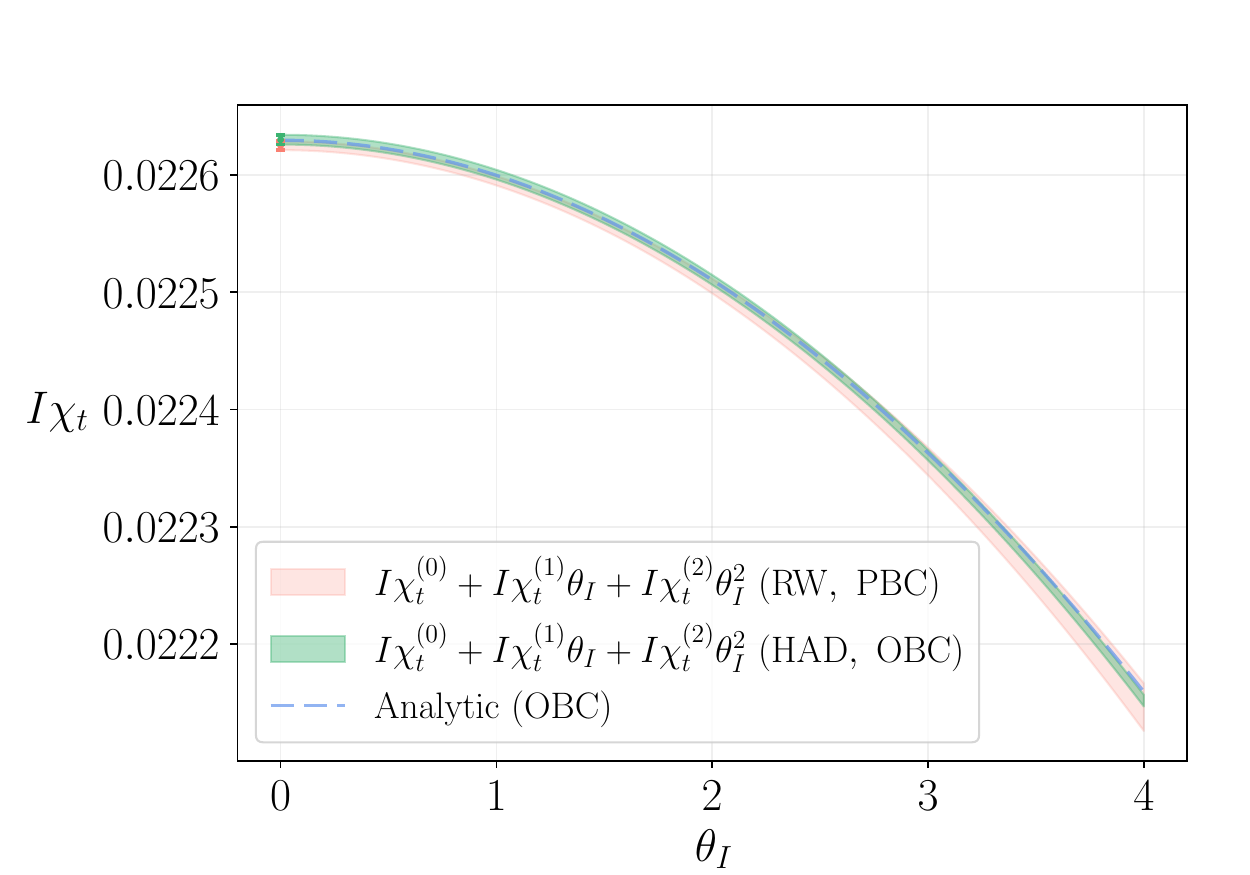}
 \caption{
   The $\theta_{I}$ dependence of $I \chi_{t}$ at $\hat{I}=5$ and $\hat{T}=100$
   computed with different methods for
   periodic boundary conditions and the standard definitions of the action and
   topological charge. Left panel: results
   (orange circles) from five different simulations with 100k uncorrelated
   configurations each, along with their fit to the functional form
   $f(\theta_{I})=a+b\theta_{I} + c\theta_{I}^2$. Right panel: results from single
   simulations with 500k uncorrelated configurations using reweighting and HAD
   (red and green points) at $\theta_{I}=0$, along with the curve
   $I\chi_{t}^{(0)}+I\chi_{t}^{(1)}\theta_{I}+I\chi_{t}^{(2)}\theta_{I}^{2}$ (light-red and thick-green
   bands) obtained from the use of truncated polynomials. The reweighting was
   performed on a simulation with periodic boundary condition, while HAD was
   used with open boundary conditions. In both panels the analytic result from
   open boundary conditions is displayed (dashed line).
 }
  \label{fig:theta-dependence-methods}
\end{figure*}

\section{The $\theta$ dependence of the spectrum}
\label{sec:theta-dependence}

The spectrum of the quantum rotor is changed by the presence of the
$\theta$-term, as seen in Eq.~(\ref{eq:spectrum continuum}). However, and
similarly to the previous section, we are interested in obtaining this
dependence explicitly from local correlators on the lattice. The obvious
challenge in this computation is the sign problem arising from the imaginary
term in the partition function of Eq.~(\ref{eq:qr-partition-fct}) when
$\theta \neq 0$.

\subsection{Overcoming the sign problem on the lattice} \label{sec:signproblem}

A common approach around the sign problem is to perform simulations at an
imaginary value of $\theta$ by defining $\theta_{I} = i\theta \in \mathbb{R}$.
This makes the integrand in Eq.~(\ref{eq:qr-partition-fct}) real and
allows us to
investigate the $\theta$-dependence of observables using standard sampling
algorithms. Analyticity around $\theta = 0$ of the observable $O$ under study is
also a required assumption, i.e.,
\begin{align}
    O(\theta) = O^{(0)} + O^{(1)}\theta + O^{(2)} \theta^2 +
    \mathcal{O}(\theta^{3}),
\end{align}
with the expansion coefficients
\begin{align}
    O^{(n)}=\frac{1}{n!}\eval{\pdv[n]{O}{\theta}}_{\theta=0}.
\end{align}
From this, we can obtain the expansion coefficient $O^{(n)}$ from analytic
continuation and fits to data. Particularly, in
Fig.~\ref{fig:theta-dependence-methods} (left), we show results of $I \chi_{t}$ at
different $\theta_{I}$ at $\hat{I}=5$, along with a fit to the functional form
$f(x) = a + b x + c x^2$ with parameters $a, b, c \in \mathbb{R}$.

This method has been used recently to study the $\theta$-dependence of the
energy spectrum of four-dimensional SU($N$) Yang--Mills theories
\cite{bonanno2024thetadependence}. One of its drawbacks, however, is that it
requires several simulations to extract the different expansion coefficients.

Alternatively, it is possible to obtain arbitrarily high expansion
coefficients by performing a single simulation where the Taylor expansion in
$\theta$ is done automatically using the algebra of truncated polynomials
\cite{alberto_ramos_2023_7970278}. From the point of view of the lattice
implementation, the idea is to define a new field, $\tilde{\phi}$, as a power
series expansion in $\theta$ up to some order $K$, i.e.,
\begin{align}
  &\tilde \phi(\theta) \equiv \tilde\phi^{(0)} + \tilde\phi^{(1)}\theta +
  \tilde\phi^{(2)}\theta^{2}+\dots+\tilde\phi^{(K)}\theta^{K}.
\end{align}
All basic operations and functions acting on these truncated polynomials are
defined to be exact at each order and, therefore, the computation of observables
preserves the correct derivatives with respect to $\theta$. For more details
about the implementation of automatic differentiation\footnote{Truncated
polynomials form the basis of the forward implementation of automatic
differentiation \cite{haro_ad}.} using truncated polynomials and applications to
stochastic processes, see Ref.~\cite{Catumba:2023ulz}.

The simplest application of truncated polynomials to extract higher order
derivatives is by using reweighting techniques. Starting from an already
existent ensemble of configurations generated at $\theta = 0$, reweighting
allows for the computation of expectation values for $\theta\neq 0$ by using the
identity
\begin{equation}
  \label{eq:rw}
  \expval{O(\phi)}_{\theta} = \frac{\expval{e^{i \theta
  Q}O(\phi)}_{\theta=0}}{\expval{e^{i \theta Q}}_{\theta=0}},
\end{equation}
where $\left< \cdots \right>_{\theta}$ denotes the expectation value with respect to
the probability distribution $p(\phi) = \frac{1}{Z(\theta)} e^{-S(\phi) + i
\theta Q}$.

To extract the Taylor expansion of $\left<O(\phi) \right>_{\theta}$  with
respect to $\theta$ up to order $K$ we replace $\theta$ with a truncated
polynomial $\tilde\theta = \sum_{k=0}^{K}\tilde{\theta}^{(k)}\theta^{k}$ where
$\tilde{\theta}^{(1)}= 1$ and $\tilde{\theta}^{(k \neq 1)}= 0$. The factor
$\exp(i \theta Q)$ and also the numerator and denominator of
Eq.~(\ref{eq:rw}) thus
become truncated polynomials. Therefore, by performing reweighting on a single
ensemble we obtain the full analytical dependence of the polynomial expansion.
This is shown in Fig.~\ref{fig:theta-dependence-methods} (right), where by
reweighting a single standard simulation at $\theta_{I} = 0$ with periodic
boundary conditions we automatically obtain $I\chi_{t}^{(k)}$; particularly, we
show $I\chi_{t}^{(0)}+ I\chi_{t}^{(1)}\theta_{I}+ I\chi_{t}^{(2)}\theta_{I}^2$
in the (light) red band, which agrees with the analytical result (dashed line)
obtained with open boundary conditions.

An alternative application of truncated polynomials consists in a modification
of sampling algorithms based on Hamiltonian dynamics inspired by numerical
stochastic perturbation theory \cite{DiRenzo:1994av,DallaBrida:2017tru}.
Particularly, we consider a modification of the HMC algorithm, which we denote as
Hamiltonian Automatic Differentiation (HAD), but this modification can be easily
applied to other sampling methods.  For the quantum rotor, since
differentiability with respect to the field is required in the HMC equations of
motion, it is necessary to use the standard discretization of the action and
topological charge in Eqs.~(\ref{eq:st-s-qr}) and (\ref{eq:st-Q-qr}).
Concretely, with periodic boundary conditions the equations of motion read
\begin{align}
    \label{eq:eom-hmc-qr}
    \dot{\phi}_{t} =& \frac{\partial H(\pi, \phi)}{\partial \pi_{t}} = \pi_{t}, \nonumber \\
    \dot{\pi}_{t} =& - \frac{\partial H(\pi,\phi)}{\partial \phi_{t}} \\
              =& - I \left[ \sin(\phi_{t}-\phi_{t-1})- \sin(\phi_{t+1}-\phi_{t})
              \right] \nonumber \\
               &+\theta_{I}\frac{1}{2\pi} \left[ \cos(\phi_{t}-\phi_{t-1}) -
               \cos(\phi_{t+1}-\phi_{t}) \right], \nonumber
\end{align}
where $t\in[0, \hat{T}-1]$, the dot denotes the derivative with respect to Monte
Carlo time, $\pi$ is the canonical conjugate momenta of $\phi$, and $H(\pi,\phi)
= \sum_{t=0}^{\hat{T}-1}\pi_{t}^2/2+ S_{\text{st}}(\phi) - \theta_{I}
Q_{\text{st}}(\phi)$ is
the Hamiltonian of the system.

To extract deritatives with respect to $\theta_{I}$, we replace it by a truncated
polynomial, $\tilde{\theta}_{I}$. Thus, $\phi$ and $\pi$ become truncated
polynomials as well, and Eq.~(\ref{eq:eom-hmc-qr}) becomes a set of differential
equations where each order involves only terms of the same or lower
order.\footnote{ The convergence of the equations of motion for each order is
    guaranteed for large values of $\hat I$. This is seen by noticing that the
    equation for each order can be rewritten as
\begin{equation*}
    \ddot\phi_{t}^{(n)} = - \pdv[2]{S}{\phi_{t}} \phi_{t}^{(n)} + \textrm{lower order terms},
\end{equation*}
and that convergence requires the positivity of the linear factor
\begin{equation*}
    \pdv[2]{S}{\phi_{t}} = \hat I \left[\cos(\phi_{t}-\phi_{t-1}) + \cos(\phi_{t+1}-\phi_{t})\right].
\end{equation*}
Since $(\phi_{t}-\phi_{t-1})\rightarrow 0$ in the continuum,
$\hat I \rightarrow\infty$, the factor $\pdv[2]{S}{\phi_{t}}$ remains positive.
For all simulations performed, the absolute value of $\phi_{t+1}-\phi_{t}$
remains in the region where convergence is guaranteed.}

The use of truncated polynomials within the solver of the equations of motion
automatizes the sampling. Therefore, after a single HMC simulation with
truncated polynomials,\footnote{In the usual HMC formalism, the Metropolis
accept-reject step is included to correct for the errors in the numerical
integration of the equations of motion. Since this discrete step is not
differentiable, it is not possible to perform it together with the expansion of the
Hamiltonian. However, it is sufficient to use a fine enough integration step size
such that the possible extrapolation to vanishing step size is below statistical
uncertainties.} we obtain a Markov chain of $N$ samples,
$\{\tilde\phi_{(i)}\}_{i=1}^{N}$, that carry the derivatives with respect to
$\theta_{I}$.  The Taylor expansion of observables is obtained by the
computation of conventional expectation values using these samples.

\begin{table}[t]
    \centering
    \begin{tabular}{ccc}
      \toprule
        $\text{Method}$ & $\chi_{t}^{(0)} \times 10^{-3}$ & $\chi_{t}^{(2)}
        \times 10^{-6}$ \\
        \midrule
        Fit & 4.5238(16) & -6.08(48)\\
        Reweighting & 4.52501(76) & -5.99(25)\\
        HAD & 4.52604(83) & -5.980(34)\\
                          \bottomrule
    \end{tabular}
    \caption{Comparison of errors between different methods to obtain the
$\theta_{I}$-dependence, namely, the quadratic fit to the results from direct
simulations at $\theta_{I}$ (Fig.~\ref{fig:theta-dependence-methods} left),
reweighting, and HAD (Fig.~\ref{fig:theta-dependence-methods} right). These results
correspond to simulations with $\hat{I}=5$ and $\hat{T}=100$, using the standard
definitions of the action and topological charge.}
    \label{tab:error-comparison-tab}
\end{table}

In Fig.~\ref{fig:theta-dependence-methods} (right), we show the curve
$I\chi_{t}^{(0)} + I\chi_{t}^{(1)}\theta_{I}+ I\chi_{t}^{(2)}\theta_{I}^2$ in
(thick) green, obtained from a single HAD simulation, and one can appreciate
that the predictions for high $\theta_I$ are more accurate than the ones
obtained by reweighting. This can be seen more transparently in
Tab.~\ref{tab:error-comparison-tab}, where we show the results for
$\chi_{I}^{(0)}$ and $\chi_{I}^{(2)}$ obtained with the three methods with an
equivalent amount of statistics. Focusing on $\chi_{t}^{(2)}$, one can see that
the reweighting result has an improved accuracy with respect to the value
obtained by simulating directly at imaginary values of $\theta$; on the other
hand, the error obtained with HAD is an order of magnitude smaller, although one
should also consider the additional computational overhead coming from the
operations of truncated polynomials.\footnote{For a more detailed cost
comparison between the reweighting and HAD methods, see
Ref.~\cite{Catumba:2023ulz}.}

\subsection{Spectrum from lattice correlators}

\begin{figure}
  \centering
 \includegraphics[width=0.49\textwidth]{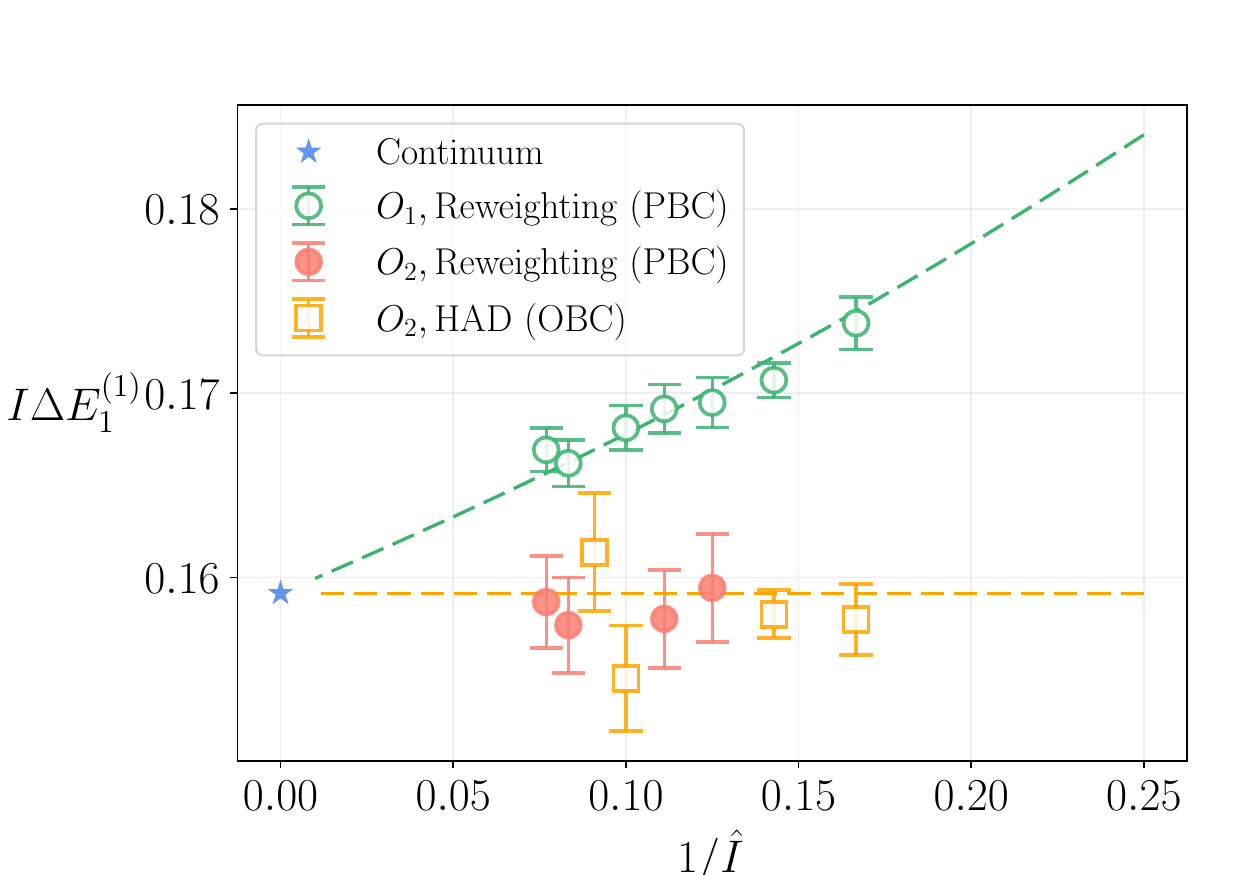}
 \caption{
    Linear $\theta$-correction to the first energy level, $\Delta E_{1}$, for a
    periodic lattice with the standard discretization of the action at different
    values of $\hat I$ with constant $T/I=20$. The results were obtained using
    the interpolating operator $O_{1}(t)=\phi_{t}$ with reweighting on a
    simulation with periodic boundary conditions (green, open circles), and
    using $O_{2}(t)=\sin(\phi_{t})$ with HAD (yellow, open squares) and
    reweighting (red, filled circles) with open and periodic boundary
    conditions, respectively. The dashed lines represent the corresponding
    analytic results obtained with open boundary conditions [see
    Eq.~(\ref{eq:standard energies})].}
  \label{fig:delta E theta}
\end{figure}

To extract the energy spectrum on the lattice, we use the spectral decomposition
of an operator $O$,
\begin{equation}
  \label{eq:spectral decomposition correlator}
    C(t) = \expval{O(t)O(0)} = \sum_k \bra 0 \hat O \ket k \bra k \hat O \ket 0
    e^{-t\Delta E_k},
\end{equation}
where $\Delta E_k = E_k - E_0$ are the energies of the system relative to the
ground state. For $t \gg 1$, the only non-negligible contribution comes from the
first energy difference,
\begin{equation}
  \label{eq:correlator large t}
    C(t) \propto \abs{\bra 1 \phi \ket 0}^2\left( e^{-\Delta E_1 t} +
    e^{-\Delta E_{-1} t} \right),
\end{equation}
where we have used the fact that $\abs{\bra 1 \phi \ket 0}^2 =  \abs{\bra{-1} \phi \ket
0}^2$, where $\bra{-1}$ denotes the energy eigenstate with eigenvalue $E_{-1}$
in Eq.~(\ref{eq:spectrum continuum}).  If we now perform a series expansion in the energy difference,
\begin{equation}
    \Delta E_n(\theta) = \Delta E_n^{(0)} + \Delta E_n^{(1)}\theta +
    \order{\theta^2},
\end{equation}
we obtain
\begin{align}
  \label{eq:obc-correlator-expansion}
    C(t) &\propto e^{- E_{1}^{(0)} t}\left[ 1 + \frac{1}{2}\theta^2 t^2(\Delta E_1^{(1)})^2
    + \mathcal{O}(\theta^{4}) \right],
\end{align}
where we have used the fact that $\Delta E_{-1}^{(1)} = - \Delta E_{1}^{(1)}$. By
computing the correlator on the lattice with the methods outlined in
Sec.~\ref{sec:signproblem}, one automatically obtains numerical results
for $C^{(0)}(t)$ and $C^{(2)}(t)$, which, from
Eq.~(\ref{eq:obc-correlator-expansion}), are proportional to
\begin{align}
C^{(0)}(t) &\propto e^{-E_{1}^{(0)}t}, \nonumber \\C^{(2)}(t) &\propto \frac{1}{2} t^2(\Delta E_{1}^{(1)})^2 e^{-E_{1}^{(0)}t}.
\end{align}
Therefore, one can obtain $\Delta E^{(1)}$ from the
large-$t$ behavior of a quadratic fit to the expression
\begin{equation}
    \label{eq:ratio correlator}
    \frac{C^{(2)}(t)}{C^{(0)}(t)} = \frac{1}{2} (\Delta E_1^{(1)})^2 t^2,
\end{equation}
which will hold for open boundary conditions. For periodic boundary conditions,
the expression to fit (see Appendix~\ref{sec:theta-correction} for a
derivation) reads
\begin{align}
  \label{eq:ratio correlator PBC}
  \frac{C^{(2)}(t)}{C^{(0)}(t)} = \frac{1}{2}(\Delta E_1^{(1)})^2\left[t^{2} +
  \frac{(T^2-2tT)}{1 + e^{- \Delta E_{1}^{(0)} (T-2t)}}\right].
\end{align}
To obtain the energy spectrum, we use two different interpolating operators,
$O_{1}(t) = \phi_{t}$ and $O_{2}(t) = \sin(\phi_{t})$. Analogously to the study
of the topological susceptibility in Sec.~\ref{sec:chi_t-from-local}, we extract
$\Delta E_{1}^{(1)}$ for increasing values of $\hat{I}$ following the line of
constant physics $T/I=20$, in order to take the continuum limit.  In particular,
in Fig.~\ref{fig:delta E theta} we plot the results for the dimensionless
quantity $I\Delta E_{1}^{(1)}$, where all data points come from simulations with
the standard action in Eq.~(\ref{eq:st-s-qr}).

Irrespective of the choice of boundary conditions, the results obtained from
local correlators match the analytical results that can be obtained with open
boundary conditions (see Appendix~\ref{sec:theta-correction}), and they have the
correct continuum limit obtained in Eq.~(\ref{eq:spectrum continuum}) using the
quantum mechanical formalism.  It is also interesting to note that, while for
this model the topological quantization depends only on the choice of boundary
conditions, the analytic $\theta$-dependence is nevertheless obtained even for
choices that do not generate topological quantization.

\section{Conclusions}

We have studied a well-known toy model of QCD to explore a recent claim to
solve the strong $CP$ problem, where it is argued that physical observables do not
depend on the $\theta$ angle of QCD because the infinite volume limit should be
taken before the summation over topological sectors, an order of limits which
makes $\theta$ disappear from the energy spectrum of the theory.  Concretely,
the authors of Refs.~\cite{ai_absence_2021,ai_consequences_2021} consider global
observables (i.e.  integrals over all spacetime), where the order of limits is
indeed relevant and subtleties about the choice of boundary conditions have to
be addressed with care. For the case of local observables, their claim is based
on the requirement of topological quantization, which is usually seen in the
limit of infinite spacetime.

In the present work, we argue that quantities of physical interest that can be
compared with experiment are intrinsically local, and therefore their dependence
on the choice of boundary conditions or global questions such as the
quantization of the global topological charge is completely irrelevant. We have
studied observables in the quantum rotor model from local correlators, and we
have shown that the requirement of topological quantization at any finite value
of the lattice spacing and volume is unimportant. In fact, we have seen that
even in cases for which the topology is not quantized, the local correlators
deliver the correct $\theta$-dependence of the observables.

Particularly we have determined the topological susceptibility and the spectrum
of the theory from local correlators with different choices of boundary
conditions (open and periodic), and also at fixed topological charge using
volume averages on very large volumes (using master field simulations).  We find
perfect agreement in the $\theta$-dependence between all the methods and the
analytical results.

For the computation of the $\theta$-dependence of the observables on the lattice
we have studied two recent proposals that might yield a computational advantage
with respect to the standard ways of alleviating the sign problem. These new
proposals involve using truncated polynomials to automatically Taylor expand
observables in $\theta$. These methods have the advantage that one can obtain the full
dependence on the first orders of $\theta$ from a single simulation at
$\theta=0$. Concretely, we have studied truncated polynomials using reweighting
techniques and the HMC algorithm. 

We conclude, that at least for the quantities of physical interest that can be
derived from local correlators (such as the susceptibility or the electric
dipole moment of the neutron in the case of QCD), the order of limits is
irrelevant, and a dependence on $\theta$ in the observables is still present.
We fail to understand why QCD respects $CP$ with such a high
accuracy.

\section*{Aknowledgements}

We especially want to thank C.  Tamarit for his patience in explaining to
us the arguments of Refs.~\cite{ai_absence_2021,ai_consequences_2021}, P. Hernandez
for extremely useful discussions, and S. Cruz-Alzaga for his contribution at
the early stages of this work. We acknowledge support from the Generalitat
Valenciana Grant No.~PROMETEO/2019/083, the European Projects
No.~H2020-MSCA-ITN-2019//860881-HIDDeN and No.~101086085-ASYMMETRY, and the National
Project No.~PID2020-113644GB-I00, as well as the technical support provided by the
Instituto de Física Corpuscular, IFIC (CSIC-UV). D.A. acknowledges support from
the Generalitat Valenciana Grants No.~ACIF/2020/011 and No.~PROMETEO/2021/083. G.C. and A.R.
acknowledge financial support from the Generalitat Valenciana Grant
No.~CIDEGENT/2019/040.  The computations were performed on the local SOM clusters,
funded by the MCIU with funding from the European Union NextGenerationEU
(PRTR-C17.I01) and Generalitat Valenciana Grant No.~ASFAE/2022/020. We also acknowledge
the computational resources provided by Finis Terrae II (CESGA), Lluis Vives
(UV), Tirant III (UV). The authors also gratefully acknowledge use of the computer
resources at Artemisa, funded by the European Union ERDF and Comunitat
Valenciana, as well as the technical support provided by the Instituto de Física
Corpuscular, IFIC (CSIC-UV).

\bigskip

\appendix

\section{Analytical lattice two-point function with open boundary conditions}
\label{sec:expl-comp-with}

Consider the generating functional for correlation functions of the topological
charge density on the lattice,
\begin{align}
  Z[J] = \int \left(\prod_{t=1}^{N} d\phi_{t}\right) \exp \bigg[ -S(\phi)
  + \sum_{t=1}^{T/a}J_{t}q_{t}  \bigg],
\end{align}
with the source $J_{t}$ and
$q_{t} = \frac{1}{2\pi}(\phi_{t+1}-\phi_{t})\bmod 2\pi$.  The explicit forms of
the discretized actions in Eqs.~(\ref{eq:cp-s-qr}) and (\ref{eq:st-s-qr})
and the topological charge in Eqs.~(\ref{eq:cp-Q-qr}) and (\ref{eq:st-Q-qr})
depend only on the combination $\phi_{t+1}-\phi_{t}$.  This suggests the change
of integration variables $\phi_{t}\rightarrow q_{t}$ defined by
\begin{equation}
  \phi_{t} = \phi_{\hat{T}-1} - \sum_{i=t}^{\hat{T}-1} q_{i},
\end{equation}
with $t\in[0, \hat{T}-1]$ and $q_{\hat{T}-1}=0$. With open boundary conditions,
every time layer decouples, leaving $\hat{T}-1$ identical independent integrals.
For the case of the classical perfect action in Eq.~(\ref{eq:cp-s-qr}), we
get
\begin{align}
  &Z[J] = 2\pi \left(-\sqrt{\frac{\pi}{2\pi}}\right)^{N}\prod_{i=1}^{N-1}
  e^{-\frac{\left(J_{i} + i\frac{\theta}{2\pi}\right)^{2}}{2 \hat{I}}}\\
  &\times\left[ \erf\left(\frac{-J_{i}-i \frac{\theta}{2\pi} - \pi
   \hat{I}}{\sqrt{2 \hat{I}}}\right) + \erf\left(\frac{J_{i}+ i \frac{\theta}{2\pi} - \pi
     \hat{I}}{\sqrt{2 \hat{I}}}\right)  \right].\nonumber
\end{align}
Differentiating the functional integral with respect to $J_{t_{1}}$ and
$J_{t_{2}}$ and setting these to zero gives us the connected two-point
correlation function of the topological charge density,
\begin{equation}
  \expval{q_{t_{1}}q_{t_{1}}}_{c} = \frac{1}{Z[0]}\eval{\frac{\delta^{2}Z[J]}{\delta J_{1}\delta J_{2}}}_{J=0}.
\end{equation}
For a finite
distance $t_{1}-t_{2}\neq 0$ the two-point function vanishes, while for
$t_{1}=t_{2}$ we get
\begin{align}
  \label{eq:twopoint}
  \expval{q_{t}q_{t}}_{c}& = \frac{1}{4\pi^{2}\hat{I}} \Bigg\{  1 + \frac{\sqrt{2
      \hat{I}}}{\sqrt{\pi}C^{2}(\theta,\hat{I})} \\
  & \quad\times\Bigg[ \frac{e^{-\frac{A^{2}( \hat{I}, -\theta)}{2\hat{I}}}}{\hat{I}} \left(
          C(\theta, \hat{I})A(
      \hat{I}, -\theta)-\hat{I}\sqrt{2/\pi} \right)\nonumber \\
  &\quad+\frac{e^{-\frac{A^{2}( \hat{I}, \theta)}{2\hat{I}}}}{\hat{I}}\left(
C(\theta, \hat{I})A( \hat{I},\theta)+\hat{I}\sqrt{2/\pi} \right) \Bigg] \Bigg\},\nonumber
\end{align}
where $A( \hat{I},\theta)\equiv \pi  \hat{I} + \frac{i\theta}{2\pi}$ and
\begin{equation}
  C(\theta,\hat{I}) = \erf\left(\frac{-A(\hat{I},\theta)}{\sqrt{2\hat{I}}}\right) +
  \erf\left(\frac{-A(\hat{I},-\theta)}{\sqrt{2\hat{I}}}\right).
\end{equation}

\begin{figure}[t]
  \centering
  \includegraphics[width=0.49\textwidth]{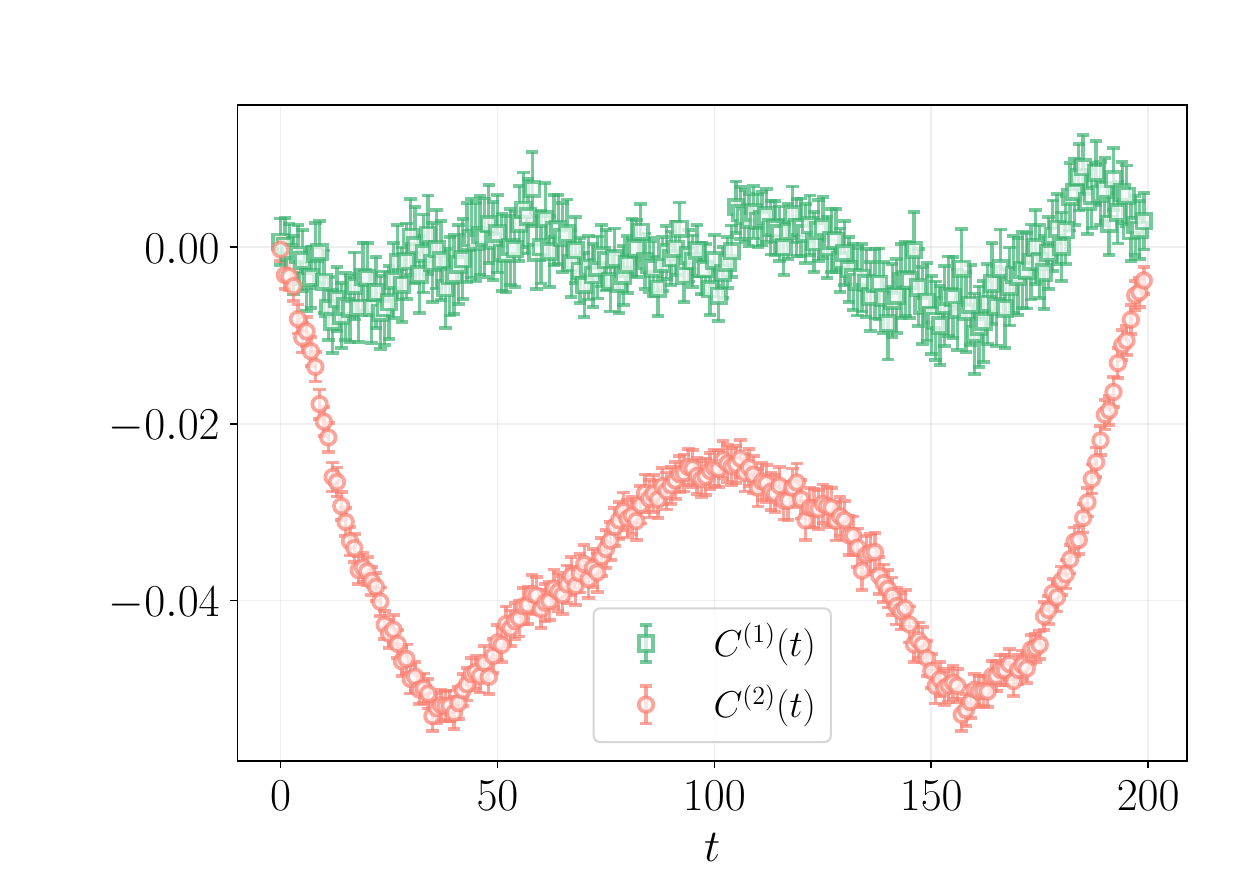}
  \caption{Linear (green squares) and quadratic (red circles) contributions to
      the two-point function $C(t)=\expval{\phi_{t}\phi_{0}}_{c}$ in the
      $\theta$-expansion from a simulation at $\hat{I}=10$ and $\hat{T}=200$
  with periodic boundary conditions and the standard action.}
  \label{fig:c1c2correlators}
\end{figure}

\noindent Eq.~(\ref{eq:twopoint}) gives the exact result of the correlation function for
any value of $\hat{I}$ and $\theta$. Additionally, the continuum value is
recovered for $\hat{I}\rightarrow\infty$. The complete lack of finite-$T$ effects is
due to the choice of boundary conditions that effectively decouple the time
layers.

The global definition of the topological susceptibility can also be computed
from the above generating functional,
\begin{align}
  \chi_{t}(\theta)^{\text{global}} &=  -\frac{1}{T}\pdv[2]{\log Z[0]}{\theta}=\frac{T-1}{T}\expval{q_{t}q_{t}}_{c},
\end{align}
recovering a similar expression up to finite-$T$ effects.

\section{Spectrum and $\theta$-dependence from the lattice}
\label{sec:theta-correction}

\begin{figure}
  \centering
  \includegraphics[width=0.49\textwidth]{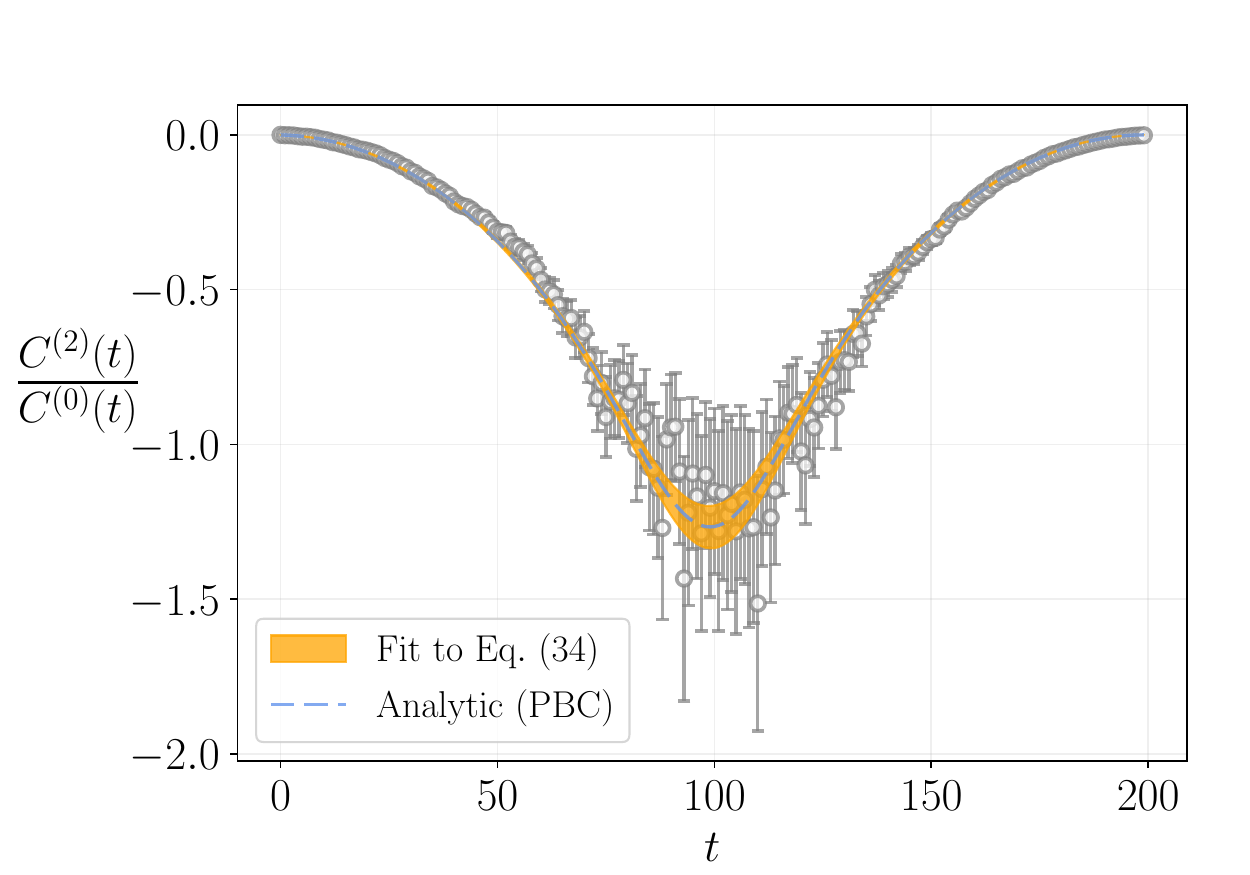}
  \caption{Ratio $C^{(2)}(t)/C^{(0)}(t)$  from a simulation at $\hat{I}=10$ and
      $\hat{T}=200$ with periodic boundary conditions using the standard action.
      The resulting fit with the functional form in Eq.~(\ref{eq:ratio
          correlator PBC}) is shown (yellow band) together with the exact
          analytical result (blue dashed line).}
  \label{fig:c2/c0 fit}
\end{figure}

\subsection{Analytical results}

The energy spectrum of the lattice theory can be computed analytically for
different choices of the action discretization. We consider a Fourier transform
of the transfer matrix \cite{bietenholz_perfect_1997} with respect to
$\psi_{t}=\phi_{t+a}-\phi_{t}$,
\begin{equation}
    A_n(\theta) \equiv e^{-aE_n(\theta)} = \int_{-\pi}^{\pi} d\psi
    \bra{\phi_{t+a}} \mathcal{T} \ket{\phi_t} e^{-in\psi}.
\end{equation}
The transfer matrix for the standard discretization of the action and
topological charge reads
\begin{align}
  \bra{\phi_{t+a}} \mathcal{T} \ket{\phi_t}_{\rm{st}} =
  \sqrt{\frac{\hat{I}}{2\pi}}\exp\bigg\{-\hat{I}(1-\cos\psi)+ i
  \frac{\theta}{2\pi}\sin\psi\bigg\},
\end{align}
and the corresponding energies are
\begin{align}
  \label{eq:standard energies}
  e^{-E_n(\theta)} = \sqrt{\frac{\hat{I}}{2\pi}} \int_{-\pi}^{\pi} d\psi\, \exp{
  -\frac{\hat{I}}{a}(1-\cos\psi)}& \nonumber\\
  \times\exp{in\psi-i\frac{\theta}{2\pi}\psi}&.
\end{align}
For the classical perfect discretization, the transfer matrix is
\begin{equation}
  \begin{aligned}
    & \bra{\phi_{t+a}} \mathcal{T} \ket{\phi_t}_{\rm{cp}} \\ &
    \begin{aligned}[t]
     \quad =\sqrt{\frac{ \hat{I}}{2\pi}}\exp\bigg\{&-\frac{ \hat{I}}{2}((\phi_ {t+1}-\phi_t)\bmod 2\pi)^2 \\
    & + i \frac{\theta}{2\pi} (\phi_ {t+1}-\phi_t)\bmod 2\pi \bigg\},
    \end{aligned}
  \end{aligned}
\end{equation}
and the energies read
\begin{align}
    e^{-E_n} = \sqrt{\frac{\hat{I}}{2\pi a}} \int_{-\pi}^{\pi} d\psi\,
    \exp{-\frac{\hat{I}}{2}\psi^2 + i\psi\left( n-\frac{\theta}{2\pi} \right)}.
\end{align}

\subsection{Lattice correlators}

The correlation functions on a finite lattice are modified due to the choice of
boundary conditions.  For a periodic lattice, each term $k$ in Eq.~(\ref{eq:spectral
decomposition correlator}) has an additional contribution $e^{-(T-t)\Delta
E_{k}}$. Its behavior at large $t$ is modified to
\begin{equation}
    \label{eq:pbc-corr-app}
  \begin{aligned}
    C(t) &= 2A e^{-\Delta E_{1}^{(1)} t}\left[ 1 + \left(\Delta E_{1}^{(0)} t \theta\right)^2
    \right]\\ &+ 2A e^{-\Delta E_{1}^{(1)} (T-t)}\left[ 1 + \left(\Delta E_{1}^{(0)} (T-t)
    \theta\right)^2 \right],
  \end{aligned}
\end{equation}
where we used $\Delta E_1^{(1)} = - \Delta E_{-1}^{(1)}$.
To extract the linear correction to the energy, we consider the ratio
\begin{align}
    \label{eq:c2-c0-app}
  \frac{C^{(2)}(t)}{C^{(0)}(t)} = \frac{1}{2}(\Delta E_1^{(1)})^2\left[t^{2} +
  \frac{(T^2-2tT)}{1 + e^{- \Delta E_1^{(0)} (T-2t)}}\right].
\end{align}

In Fig.~\ref{fig:c1c2correlators}, the two first contributions to the two-point
correlator $\expval{\phi_{t}\phi_{0}}_{c}$ in the $\theta$ expansion are shown
for a periodic lattice for a simulation with the standard discretization of the
action using periodic boundary conditions.  The linear term is compatible with
zero, as expected from Eq.~(\ref{eq:pbc-corr-app}).  The ratio in
Eq.~(\ref{eq:c2-c0-app}) is plotted in Fig.~\ref{fig:c2/c0 fit}. The
fit is shown to match the exact value for this discretization, obtained with the
functional form in Eq.~(\ref{eq:c2-c0-app}) and the energy values from
Eq.~(\ref{eq:standard energies}).

\newpage

\section{Winding transformation}
\label{app:winding-trafo}

We define the winding transformation on a quantum rotor configuration as
\begin{align}
    \mathcal{W}^{\pm}:\, \phi_{t} \to  \phi_{t}^{\mathcal{W}^{\pm}} = \phi_{t}
    \pm  2\pi t / \hat{T}, 
\end{align}
for $t \in[0, \hat{T}-1]$. Assuming a smooth enough configuration $\phi$ such
that $(\phi_{t+1}-\phi_{t} \bmod2\pi) \pm 2\pi / \hat{T} \in [-\pi,\pi)$ ---which is the
case close enough to the continuum---one can show that the winding transformation
changes its classical topological charge in one unit, i.e.,
\begin{align}
    Q_{\text{cp}}(\phi^{\mathcal{W}^{\pm}}) = Q_{\text{cp}}(\phi) \pm 1.
\end{align}
The change in the classical perfect action after the winding transformation is
\begin{align}
    \Delta S_{\text{cp}}^{\mathcal{W}^{\pm}} \equiv&\,
    S_{\text{cp}}(\phi^{\mathcal{W}^{\pm}})
    - S_{\text{cp}}(\phi) \nonumber \\
    =&\, 2 \pi^2 \left( 1 \pm  2Q_{\text{cp}}(\phi) \right)
    \frac{\hat{I}}{\hat{T}},
\end{align}
and it can be made small by increasing the time extent $\hat{T}$ of the lattice.
One can use this transformation to propose new samples in a Metropolis
algorithm, to be accepted with probability
\begin{align}
    p_{\text{acc}}(\phi^{\mathcal{W}^{\pm}} \mid \phi) = \min \left\{ 1, \exp(-\Delta
    S_{\text{cp}}^{\mathcal{W}^{\pm}}) \right\}.
\end{align}
Note that one must use this update step in combination with other sampling
algorithms to ensure ergodicity \cite{Albandea_2021}.

\newpage

\bibliography{sn-bibliography.bib}

\end{document}